\documentclass{llncs}

\usepackage{times}
\usepackage{latexsym}
\usepackage{amsmath}
\usepackage{amssymb}
\usepackage{graphicx}

\usepackage[utf8]{inputenc}
\usepackage{paralist}
\usepackage{xspace}
\usepackage{subfigure}
\usepackage{cite}
\usepackage{wrapfig}

\newcommand{\eg}{e.\,g.,\xspace}
\newcommand{\ie}{i.~e.,\xspace}
\newcommand{\cf}{cf.\xspace}

\newcommand{\flickr}{Flickr\xspace}
\newcommand{\twitter}{Twitter\xspace}
\newcommand{\bibs}{BibSonomy\xspace}
\newcommand{\et}{et al.\xspace}

\newcommand{\facebook}{Facebook\xspace}

\newcommand{\conferator}{\textsc{Conferator}\xspace}
\newcommand{\mygroup}{\textsc{MyGroup}\xspace}
\newcommand{\ubicon}{\textsc{Ubicon}\xspace}
\newcommand{\VIKAMINE}{\textsc{Vikamine}\xspace}
\newcommand{\CMD}{\textsc{Comodo}\xspace}
\newcommand{\GPG}{\textsc{GP-Growth}\xspace}
\newcommand{\pagerank}{PageRank\xspace}

\newcommand{\work}{article\xspace}
\newcommand{\summary}{work\xspace}

\begin{document}
\mainmatter

\title{Data Mining on Social Interaction Networks}

\titlerunning{Data Mining on Social Interaction Networks}

\author{%
   Martin Atzmueller
}

\institute{%
    University of Kassel, Knowledge and Data Engineering Group,\\
    Wilhelmsh\"oher Allee 73, 34121 Kassel, Germany\\
    \email{atzmueller@cs.uni-kassel.de}
}

\maketitle

%%%%%%%%%%%%%%%%%%%%%%%%%% Abstract %%%%%%%%%%%%%%%%%%%%%%%%%%%%%%%%%%%%
\section*{Abstract}

Social media and social networks have already woven themselves into the very fabric of everyday life. This results in a dramatic increase of social data capturing various relations between the users and their associated artifacts.
In such settings, data mining and analysis plays a central role: Predictive data mining targets the acquisition and learning of specific models in order to support the users, \eg for classification or inference of parameters for future cases.
Furthermore, descriptive data mining aims at obtaining patterns which summarize and characterize the data.

From an application perspective, there is a variety of computational social systems -- with an increasing use of mobile and ubiquitous technologies.
The various direct and indirect interactions between the users in the online networks as well as the real-world human interactions using ubiquitous devices can then be represented using social interaction networks.

In this \work, we consider social interaction networks from a data mining perspective -- also with a special focus on real-world face-to-face contact networks: We combine data mining and social network analysis techniques for examining the networks in order to improve our understanding of the data, the modeled behavior, and its underlying emergent processes. Furthermore, we adapt, extend and apply known predictive data mining algorithms on social interaction networks. Additionally, we present novel methods for descriptive data mining for uncovering and extracting relations and patterns for hypothesis generation and exploration by the user, in order to provide characteristic information about the data and networks.
The presented approaches and methods aim at extracting valuable knowledge for enhancing the understanding of the respective data, and for supporting the users of the respective systems. We consider data from several social systems, like the social bookmarking system \bibs, the social resource sharing system flickr, and ubiquitous social systems: Specifically, we focus on data from the social conference guidance system \emph{Conferator} and the social group interaction system \emph{MyGroup}.

This \summary first gives a short introduction into social interaction networks, before we describe several analysis results in the context of online social networks, as well as real-world face-to-face contact networks. Next, we present predictive data mining methods making use of the social interactions, \ie for localization, recommendation and link prediction. After that, we present novel descriptive data mining methods for mining communities and patterns on social interaction networks.

\clearpage
%%%%%%%%%%%%%%%%%%%%%%%%%%%%%%%%%%%%%%%%%%%%%%%%%%%%%%%%%%%%%%%%%%%%%%%%%%%%%%%

%%%%%%%%%%%%%%%%%%%%%%%%%%%%%%%%%%%%%%%%%%%%%%%%%%%%%%%%%%%%%%%%
%%% Section: Introduction
%%%%%%%%%%%%%%%%%%%%%%%%%%%%%%%%%%%%%%%%%%%%%%%%%%%%%%%%%%%%%%%%
\section{Introduction}

The emergence of new social systems and organizational social applications has created a number of novel social and ubiquitous environments. By interacting with such systems, the user is leaving traces within the different databases and log files, \eg by updating the user's status via \twitter, commenting an image in \flickr, copying a post in \bibs, or connecting to other users via ubiquitous devices.

Ultimately, each type of such a trace gives rise to a corresponding network of user relatedness, where users are connected if they interacted either explicitly (\eg by a direct encounter via an RFID tag, or by establishing ``friendship'' in an online social network) or implicitly (\eg by visiting a user's profile page). We consider a link within such a network as evidence for user relatedness and interaction and call it accordingly \emph{social interaction network}. This connects but also transcends private and business applications featuring a range of different types of networks, organizational contexts and corresponding interactions, \eg networks that involve spatial proximity relations like co-location or face-to-face proximity. In this context, physical devices, \eg mobile phones or RFID devices, can help to link relations in the digital domain to relations in physical and/or social space, and vice-versa. With the growth and availability of the collected data, there is also an increasing interest in the analysis of such social interaction networks.

This \work considers data mining in social interaction networks, specifically human behavioral (offline) networks, that is, networks of face-to-face proximity, \eg during a conversation. We call these networks \emph{face-to-face contact networks} in the following.
We include social (online) networks contructed from subject-centric or object-centric sociality~\cite{KC:97}, considering data from social bookmarking systems such as \bibs\footnote{\url{http://www.bibsonomy.org}}~\cite{Bibs:VLDB10} and resource sharing systems like \flickr\footnote{\url{http://www.flickr.com}}, but especially focus on real-world face-to-face contact networks.

The capture of human face-to-face contacts in social interaction networks and the analysis of both offline and online data is receiving increasing interest. While there has been foundational work on the analysis of face-to-face contact networks, \eg~\cite{CVBCPV:10,BCSBA:10,SVBCCIRPKBV:11}, data mining on those networks is still a rather new field of research, for which we provide novel analyses resulting in new insights into their structure and relations. In addition, we adapt, extend and apply known data mining methods to the collected social interaction networks, especially the face-to-face contact networks. Furthermore, we propose novel methods and approaches for describing and characterizing the networks and properties of their nodes, respectively. Integrated into different systems~\cite{AL:12a} and applications~\cite{ABDHJMMSS:11,ABDHKMMMSS:12}, the techniques and methods are also deployed in a practical real-world setting. 

This \summary, an adapted and substantially extended revision of~\cite{Atzmueller:12c}, provides an overview on those previously published articles~\cite{ABDHJMMSS:11,ADHMS:12,MSAS:12,AM:11a,LBA:12,AL:13,SDAHS:11,SAS:12,MAS:11,MABHS:11,51seitenPaper}
as grouped together in the author's habilitation thesis~\cite{Atzmueller:2013}.

The face-to-face contact networks in our context are acquired using the \conferator\footnote{\url{http://www.conferator.org}}~\cite{ABDHJMMSS:11} and \mygroup\footnote{\url{http://ubicon.eu/about/mygroup}}~\cite{ABDHKMMMSS:12} systems.
\conferator, a social conference guidance system, and \mygroup, a social workgroup support system, are ubiquitous social systems for enhancing social interactions in the context of conferences and working groups, respectively. They are built on top of the RFID-based proximity sensing hardware developed by the SocioPatterns\footnote{\url{http://www.sociopatterns.org}} collaboration, and on top of the \ubicon software platform\footnote{\url{http://www.ubicon.eu}}~\cite{ABDHKMMMSS:12}.
Both systems allow the collection of real-world networks of human face-to-face interactions -- as behavioral networks -- as well as the utilization of additional online social networks. Since these are our own systems, we were able to comprehensively perform our experiments using all the available data.

Data mining provides approaches for the identification and discovery of non-trivial patterns and models hidden in large collections of data~\cite{FPS:96b,HK:06,Atzmueller:11a}.
While there exist several process models for data mining~\cite{LM:06}, a prominent model is given by the CRISP-DM process~\cite{WH:00,CRISPDM} -- an industry standard for data mining.
CRISP-DM consists of six phases:
\begin{inparaenum}[(1)]
\item \emph{Business Understanding}
\item \emph{Data Understanding}
\item \emph{Data Preparation}
\item \emph{Modeling}
\item \emph{Evaluation}, and
\item \emph{Deployment}.
\end{inparaenum}

We combine data mining and social network analysis techniques for analyzing social interaction networks in order to improve our \emph{understanding} of the data, the modeled behavior, and its underlying processes, in Section~\ref{sec:analysis}. This corresponds to the business and data understanding phases in the CRISP-DM process. Furthermore, for the analysis itself we can also apply techniques for explorative analysis discussed below.

Next, we focus on elements of the \emph{modeling} phase, \ie the core data mining step in Sections~\ref{sec:predictive}-\ref{sec:descriptive}: The applied data mining methods can be divided into descriptive and predictive methods~\cite{HK:06}: While descriptive methods are used for summarizing the data, for identifying hidden information in the form of patterns, and for exploration, predictive methods are used for constructing models for inferring future properties given new data, \eg for classification: We adapt and extend known predictive data mining algorithms on social interaction networks for supporting the users in typical tasks such as recommendation and localization in the context of the mentioned systems. Additionally, we present novel methods for descriptive data mining for uncovering and extracting relations and patterns for hypothesis generation and exploration by the user, in order to provide characteristic information about the data and networks.

The methods and criteria in the \emph{evaluation} phase depend on the applied model type, and are therefore specific for a data mining technique. Therefore, we discuss these in the respective subsections of Sections~\ref{sec:predictive}-\ref{sec:descriptive}. In addition, Section~\ref{sec:analysis:structural} describes a novel community assessment technique, while Section~\ref{sec:descriptive:exploratory} summarizes interactive techniques for the evaluation of mined patterns in social interaction networks. Finally, the \emph{deployment} phase is tackled by the integration of the methods into practical applications, \eg into \conferator and \mygroup as described in Section~\ref{sec:basics:conferator}. Furthermore, especially the descriptive data mining methods have been integrated into the pattern mining and analytics system \VIKAMINE~\cite{AL:12a,Atzmueller:12c}.

\subsubsection{Related Work.}
Overall, data mining in the context of social interaction networks concerns core elements of data mining and knowledge discovery itself, \eg~\cite{HK:06}, but also includes techniques from social network analysis, \eg~\cite{WF:94,SSMRBDCPG:09}, as well as mining social media, \eg~\cite{TL:10,Russell:11}, complex network analytics~\cite{BKMRRSTW:00,MMGDB:07,Newman:03,AB:02,Strogatz:01}, and mining the ubiquitous web~\cite{ZLY:02,HS:10,Sheth:10}.

Specifically, community detection~\cite{Newman:04,LF:09}, analysis of roles~\cite{STE:07a,STE:07b}, contact patterns~\cite{CVBCPV:10,DBLP:journals/corr/abs-1006-1260}, localization~\cite{NLLP:04,HVBW:00,WCI:07}, recommendations~\cite{SS:09,BHJNRSSS:12,BGHJ:11,JMHSS:08}, link prediction~\cite{LK:03,Barabasi:02,WPSGB:11}, descriptive pattern mining~\cite{Webb:08,KLW:09}, exceptional model mining~\cite{LFK:08,DKFL:10,KFL:11}, but also techniques for reality mining, \eg~\cite{EP:06,Mitchell:09} are prominent topics in this area. Especially for face-to-face contact networks, Cattuto et colleagues~\cite{CVBCPV:10,DBLP:journals/corr/abs-1006-1260,BCCPBV:08,BCSBA:10} provide an overview on social dynamics in those networks. Their experiments include applications and analysis at conferences~\cite{ASCBCB:09,BCSBA:10,SCBBA:10,DBLP:journals/corr/abs-1006-1260}, schools~\cite{SVBCIPQBRLV:11}, and in epidemiology~\cite{SVBCCIRPKBV:11,SBBBBBCCKMV:12}.
In the following sections of this \work, we will discuss related work in the respective sections in more detail.

\subsubsection{Structure of this Article.} 
The remainder of this \summary is structured as follows: Section~\ref{sec:basics} outlines basics of social interaction networks, real-world face-to-face contact networks, and describes the \conferator and \mygroup systems. Section~\ref{sec:analysis} summarizes several analysis directions concerning interrelations in social interaction networks, communities and roles in human face-to-face contact networks, and structure and dynamics of interactions of conference participants.
Next, Section~\ref{sec:predictive} discusses adaptations, extensions, and applications of predictive methods in social interaction networks.
After that, Section~\ref{sec:descriptive} presents novel efficient descriptive methods for community mining and exceptional model mining, as well as an exploratory pattern mining approach on social interaction networks. 
Finally, Section~\ref{sec:conclusions} concludes with a summary and outlook.

%%%%%%%%%%%%%%%%%%%%%%%%%%%%%%%%%%%%%%%%%%%%%%%%%%%%%%%%%%%%%%%%
%%% Section: Basics
%%%%%%%%%%%%%%%%%%%%%%%%%%%%%%%%%%%%%%%%%%%%%%%%%%%%%%%%%%%%%%%%
\section{Basics of Social Interaction Networks}\label{sec:basics}

In this section, we focus on basics of social interaction networks. We first provide an introduction to social interaction networks. Next, we focus on human face-to-face contact networks. After that, we present the \conferator and \mygroup systems.

\subsection{A Brief Introduction to Social Interaction Networks}\label{sec:basics:social}

With the rise of social software and social media, a wealth of user-generated data and user interactions is being created in online social networks~\cite{Heidemann:10} -- as a technical platform. Often, these are also called (online) social network services, \cf~\cite{AHKMJ:07}.
In the following, we adopt an intuitive definition of social media, regarding it as online systems and services in the ubiquitous web, which create and provide social data generated by human interaction and communication~\cite{Atzmueller:12c}.

A \emph{social network} is a core concept of social network analysis~\cite{Scott:11,WF:94}:
According to Wassermann and Faust~\cite{WF:94}, a social network is a social structure consisting of a set of actors (such as individuals or organizations) and a set of dyadic ties between these actors. There are various types of ties such as, for example, friendship, kinship, or organizational position. 
Usually, these relations are modeled as a graph, with the actors as nodes, and the ties as edges connecting the nodes.

In this \work, we focus on \emph{social interaction networks}~\cite{MBSH:10,MABHS:11,MASH:13,51seitenPaper}, \ie user-related social networks in social media capturing social relations inherent in social interactions, social activities and other social phenomena which act as proxies for social user-relatedness. Therefore, according to the categorization of Wassermann and Faust~\cite[p. 37 ff.]{WF:94} social interaction networks focus on \emph{interaction} relations between \emph{people} as the corresponding actors.
This also includes interaction data from sensors and mobile devices, as long as the data is created by real users.
Consider, for example, users who connect their mobile phones via Bluetooth or NFC (Near Field Communication), look at similar pictures in \flickr, talk about similar topics in \twitter, or explicitly establish ``contacts'' within certain applications. Furthermore, we consider real-world contacts as determined by other ubiquitous applications, based on principles of ubiquitous computing~\cite{Weiser:91,Weiser:93} or the ubiquitous web~\cite{ZLY:02,HS:10,Sheth:10}.

Networks in ubiquitous and social applications, for example, in RFID-based systems, can be derived according to the detected contacts between the respective RFID tags: A tie between two actors can be derived, for example, if their assigned RFID tags were in contact, possibly weighted with the number of contacts, or their durations.
Such social interation networks are often observed during certain events, for example, during conferences~\cite{ABDHJMMSS:11,MSAS:12,XCWCZYWZ:11,ZCWWWT:12}, at work~\cite{ABDHKMMMSS:12,MAS:11}, or other group-based activities.
In Section~\ref{sec:basics:f2f}, we specifically discuss human face-to-face contact networks. Furthermore, we summarize according applications in Section~\ref{sec:basics:conferator}.

Overall, we consider social interactions in an online and offline context, that is, connections and relations in online systems as well as real-world face-to-face contacts. Furthermore, we also consider social relations implemented using specific resources or artifacts, according to the principle of object-centric sociality~\cite{KC:97}, where objects of a specific actor, \eg resources, mediate connections to other actors. This is especially relevant in the scope of social bookmarking and social resource sharing systems.
Examples are given by the ``Favorite'' relation in \flickr~\cite{MKSGDB:08}, or resource - resource relations, \eg by a common set of tagged images. This also transcends to further collaborative systems, for example, data of versioning systems like CVS~\cite{MAS:11}.

Typical analytical questions concern the analysis of key actors, roles and communities, ranging from different measures of centrality, \cf~\cite{BE:04} to the exploration of topological graph properties, \eg~\cite{Gaertler:04} or structural neighborhood similarities, \eg~\cite{Lerner:05}.
The analysis of communities intuitively considers densely connected subgroups of actors, represented as nodes in the social network, \eg~\cite{GN:02,Newman2006fcs,FC:07,LF:09}.
The detection and characterization of communities is a typical descriptive data mining task~\cite{AM:11a}, as well as the description of patterns given properties of the nodes, \eg roles, tagging information, or common geo-location.
Predictive approaches include, for example, link prediction techniques, as well as recommendation, or utilizing the network for improving localization methods.
Both descriptive methods and predictive modeling approaches will be discussed in Sections~\ref{sec:predictive} and~\ref{sec:descriptive}, respectively, below.

\clearpage

\subsection{Real-World Interactions: Face-to-Face Contact Networks}\label{sec:basics:f2f}

\begin{wrapfigure}{R}{0.45\columnwidth}
\vspace{-1.2cm}
\centering
  \includegraphics[width=0.44\columnwidth]{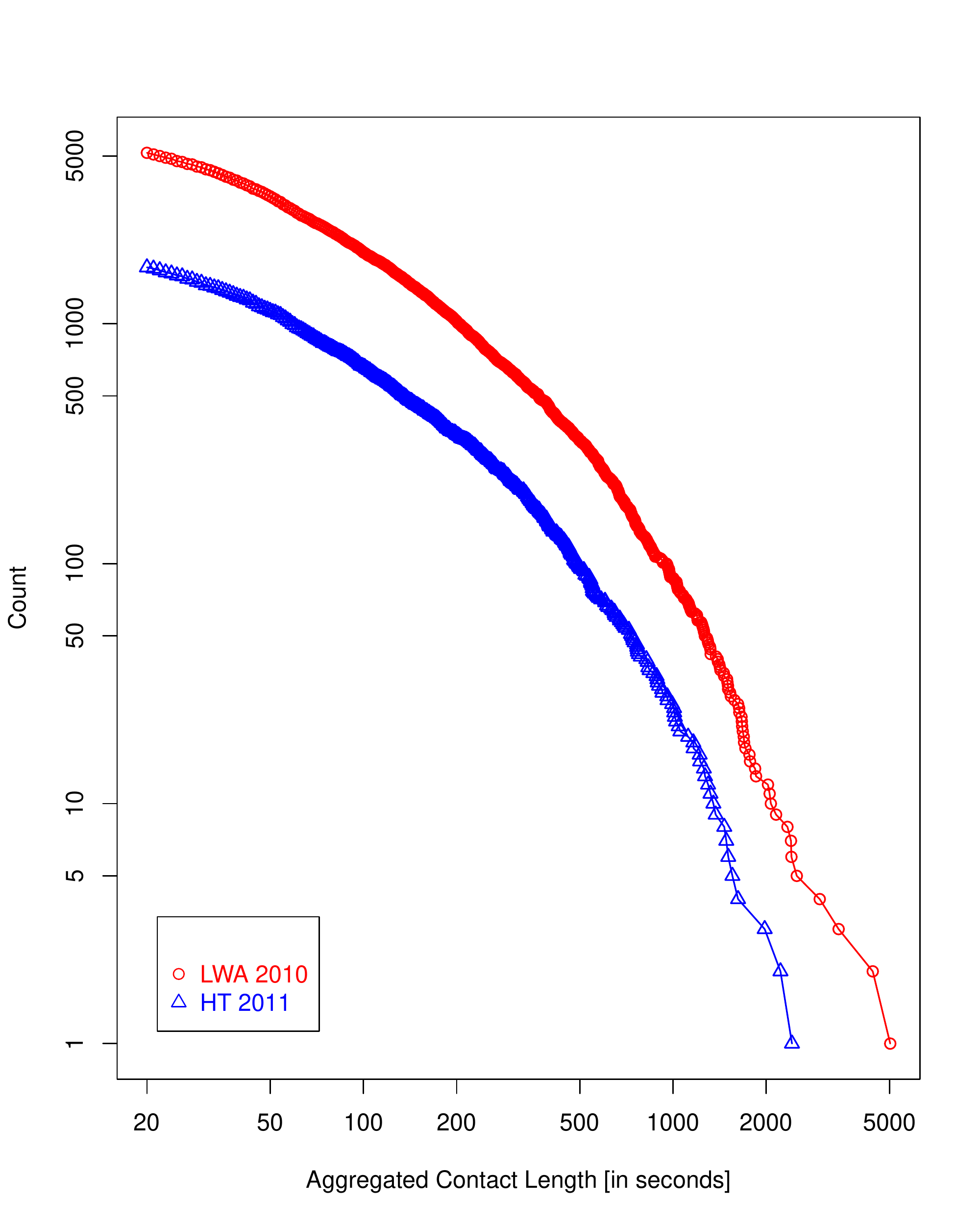}    
  \caption{Example~\cite{SAS:12} of a typical cumulated contact length distribution of human face-to-face contacts, collected at two conferences (LWA 2010 and HT 2011). The $x$-axis displays the minimum length of a contact (in seconds), the $y$-axis the number of contacts having at least this contact length.}
  \label{fig:contactlengthdistribution}
\vspace{-.5cm}
\end{wrapfigure}

In contrast to online interactions, face-to-face contacts represent interactions between actors in the real-world. Consider conference contact networks, for example: Using the SocioPatterns RFID tags and the \conferator system described below, face-to-face contacts can be measured. When the RFID tags are worn on the chest of the conference participants, tag-to-tag proximity is a proxy for face-to-face communication, since the range of the signals is approximately 1.5 meters if not blocked by the human body.
For detecting a contact, we apply the technique described in~\cite{SCBBA:10}, which is based on a typical conversational setting: A face-to-face contact is observed when the duration of the contact is at least 20 seconds. A contact ends when the two corresponding proximity tags do not detect each other for more than 60 seconds.

Face-to-face contact networks are a special sort of social interaction networks.
Using face-to-face contacts between pairs of actors, we can derive a face-to-face contact networks. The networks are constructed such that links connect actors, if these were in face-to-face contact, possible weighted with the contact count or the (normalized) contact duraction. Other networks that indicate real-world interactions, \eg by co-authorship or frequent co-location, are obtained by considering co-visited talks, posters, or co-authored papers. Networks based on co-location can be constructed, for example, based on the counts or (normalized) durations that pairs of participants were observed in the same location, for example, at room-level.
Figure~\ref{fig:contactlengthdistribution} shows an example of the typical contact distribution that we observe in conference face-to-face contact networks: Confirming previous findings, \eg by the Sociopatterns collaboration~\cite{DBLP:journals/corr/abs-1006-1260},  most of the contacts take less than one minute and the contact durations of both conferences show a long-tailed distribution.

In contrast to the work presented by the SocioPatterns collaboration discussed above, we aim at a larger research focus concerning data mining and social network analysis: We examine social dynamics in conferences and the contexts of working groups. Additionally, we especially investigate communities, and roles of actors in these networks grounded using background information. Furthermore, we do not primarily approach the \emph{modeling} of the networks phenomena from a social network analysis perspective, but focus on the analysis, and especially on approaches and methods for mining and extracting useful models and patterns from the data.

\subsection{\conferator and \mygroup: Enhancing Social Interactions}\label{sec:basics:conferator}

\conferator~\cite{ABDHJMMSS:11} and \mygroup~\cite{ABDHKMMMSS:12} are ubiquitous social systems for enhancing social interactions: \conferator is a social conference guidance system for efficiently managing face-to-face contacts at a conference, and collectively building a personalized conference program. \mygroup is a similar system in the context of working groups that aims to enhance interactions and knowledge exchange between the individual team members.
Both systems are built on top of the RFID-based proximity sensing system developed by the SocioPatterns\footnote{\url{http://www.sociopatterns.org}} collaboration. The applied RFID tags allow the coupling of real world (offline) data, \ie face-to-face contacts, with the online social world, \eg given by online interactions within the system or in linked online social networks. In particular, these RFID proximity tags can collect face-to-face contacts. This allows for highly personalized profiles in the systems which can be applied, \eg for community mining, recommendations, or for improving the localization.

In the following, we first give an overview on the systems and the applied technical platform \ubicon. After that, we summarize the most important features of the systems.

\subsubsection{Overview.}

\conferator~\cite{ABDHJMMSS:11} is a social and ubiquitous conference guidance system, aiming at supporting conference participants during conference planning, attendance and their post-conference activities. It features the ability to observe and to manage social and face-to-face contacts during the conference and provides a number of features for supporting social networking.

In a similar context, the conference navigator by Brusilovsky~\cite{WBP:10,FB:07} allows attendees of a conference to organize the conference schedule. However, it is not connected to the real live activity of the user during the conference.
Hui et al.~\cite{HCSGCD:05} describe an application using Bluetooth-based modules for collecting mobility patterns of conference participants. Furthermore, Eagle and Pentland~\cite{EP:06} present an approach for collecting proximity and location information using Bluetooth-enabled mobile phones, and analyze the obtained networks.
Similarly, the Find-And-Connect~\cite{XCWCZYWZ:11} system utilizes bluetooth and passive RFID for obtaining locations of participants, and infers \emph{encounters} based on the co-location of participants as a proxy for contacts between participants. However, no direct face-to-face contacts are measured. 
In contrast to these systems, \conferator is able to collect the real-world face-to-face contacts using the SocioPattern RFID tags described above.
The setup requires a number of RFID readers at fixed positions in the target area; the participants are then equipped with RFID tags.
The technology allows for the localization of tags and for detecting tag-to-tag proximity. When the tags are worn on the chest, tag-to-tag proximity is a proxy for a face-to-face contact, since the range of the signals is approximately 1.5 meters if not blocked by the human body.
For more details, we refer to Barrat et al.~\cite{CVBCPV:10}.

Utilizing RFID proximity tags, social interaction data can be collected at a much more accurate level than, \eg based on co-location information~\cite{EP:06,XCWCZYWZ:11}. From the data, we can derive social interaction networks, apply these for mining the collected data, and put the discovered patterns and learned models to use by the system. 
With \conferator, we provide a broad application spectrum and features: Compared to previous RFID-based approaches, we increased the precision of the localization component and linked together tag information and the conference schedule. Furthermore, we implemented a light-weight integration with \bibs, and added connectors to other social systems used by participants, \eg \facebook, \twitter, Xing or LinkedIn. This provides the basis for new insights into the behavior of all participants concerning their real-world (offline) and online social interactions.
\conferator has been successfully applied at LWA 2010\footnote{\url{http://www.kde.cs.uni-kassel.de/conf/lwa10}}~\cite{ADHMS:12}, LWA 2011\footnote{\url{http://lwa2011.dke-research.de}}, and LWA 2012\footnote{\url{http://lwa2012.cs.tu-dortmund.de}} -- conferences for special interest groups of the German Computer Science Society (GI), at the Hypertext 2011\footnote{\url{http://ht2011.org}} conference~\cite{MAS:11}, and at a technology day of the \textsc{Venus} project.\footnote{\url{http://www.uni-kassel.de/eecs/iteg/venus}}

\mygroup aims at supporting members of working groups. It employs the same technology as \conferator for localizing the members and for monitoring their social contacts. Additionally it provides profile information including links to (external) social software, \eg BibSonomy~\cite{Bibs:VLDB10}, \twitter, Facebook, or XING.
\mygroup has been applied at a number of different events: Since December 2010, it is being continuously applied by the Knowledge and Data Engineering (KDE) group at the University of Kassel, and is currently being extended towards a larger research cluster.
In addition, \mygroup has also been utilized at a large student party,\footnote{\url{http://wintersause.de}} for supporting organizational processes, at the First International Changemaker-Camp\footnote{\url{http://www.knowmads.nl}} at the University of Kassel for profiling group processes, and at a CodeCamp for supporting software development in the context of the \textsc{Venus} project.

\begin{figure}[htb]
  \centering
  \includegraphics[trim=0 3.5cm 0 0,clip,width=.8\columnwidth]{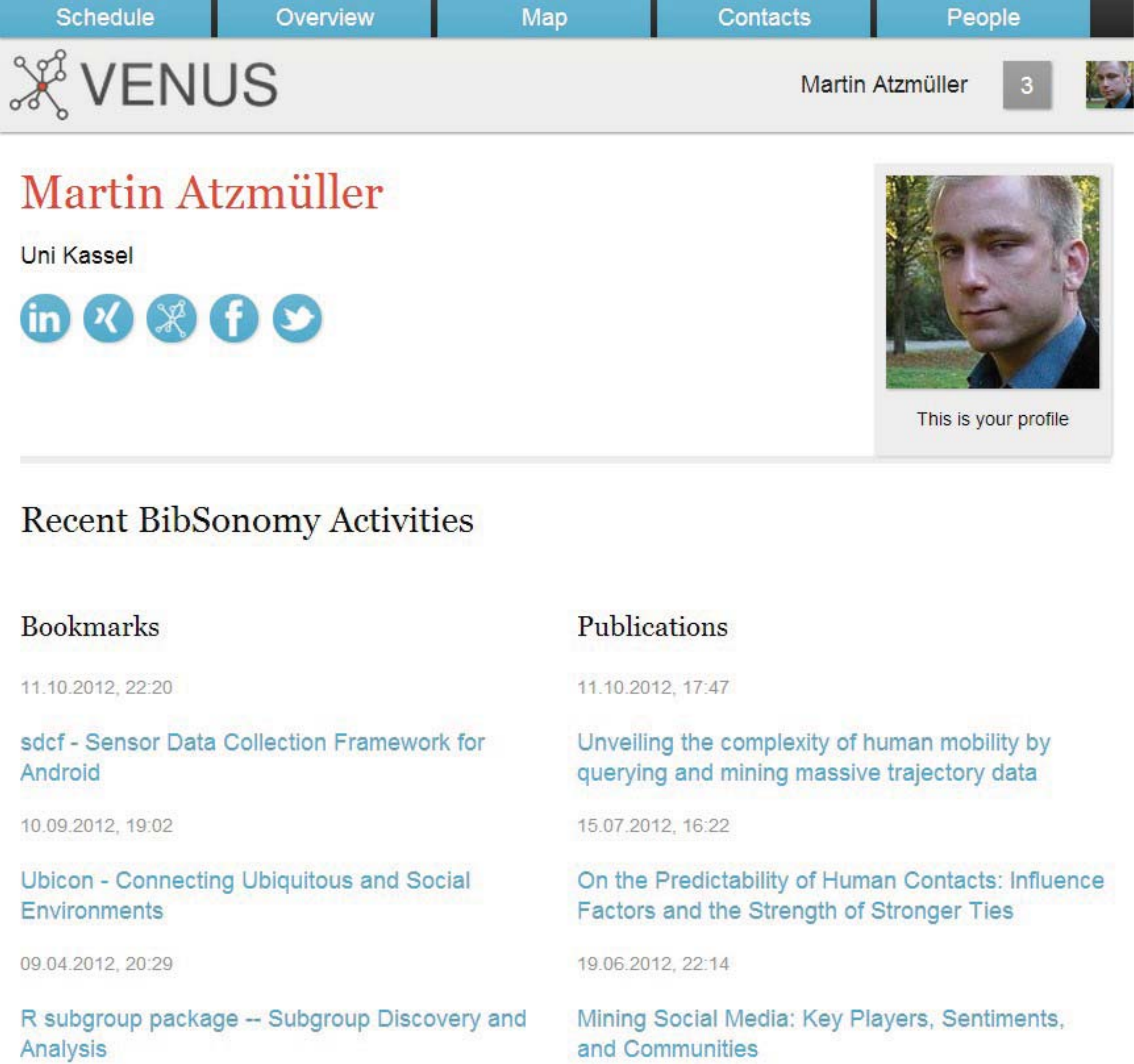}
  \caption{A screenshot~\cite{ABDHKMMMSS:12} of a \conferator
    user profile, with contact information and latest \bibs activities, \ie publication and bookmark posts. Further profile information (not shown) includes, for example, a personalized social tag cloud and location and contact context information.}
  \label{fig:peerRadar}
\end{figure}

\conferator and \mygroup have been implemented using the \ubicon platform -- a platform for enhancing ubiquitous and social networking. 
From a technical point of view, the \ubicon platform consists of the application logic, components for privacy management and database management, a (customizable) set of data processors that process the incoming (raw) data, a set of data processors for more subsequent sophisticated processing, and a storage architecture based on a MySQL database.\footnote{\url{http://mysql.com}} The set of data processors include, \eg the localization component for determining the location of RFID tags.
The system is implemented with a model-view-controller pattern using the Spring framework.\footnote{\url{http://springsource.org}} \ubicon can be deployed using a standard servlet container, \eg Apache Tomcat.\footnote{\url{http://tomcat.apache.org}}

\begin{figure}
  \centering
  \includegraphics[width=.78\columnwidth]{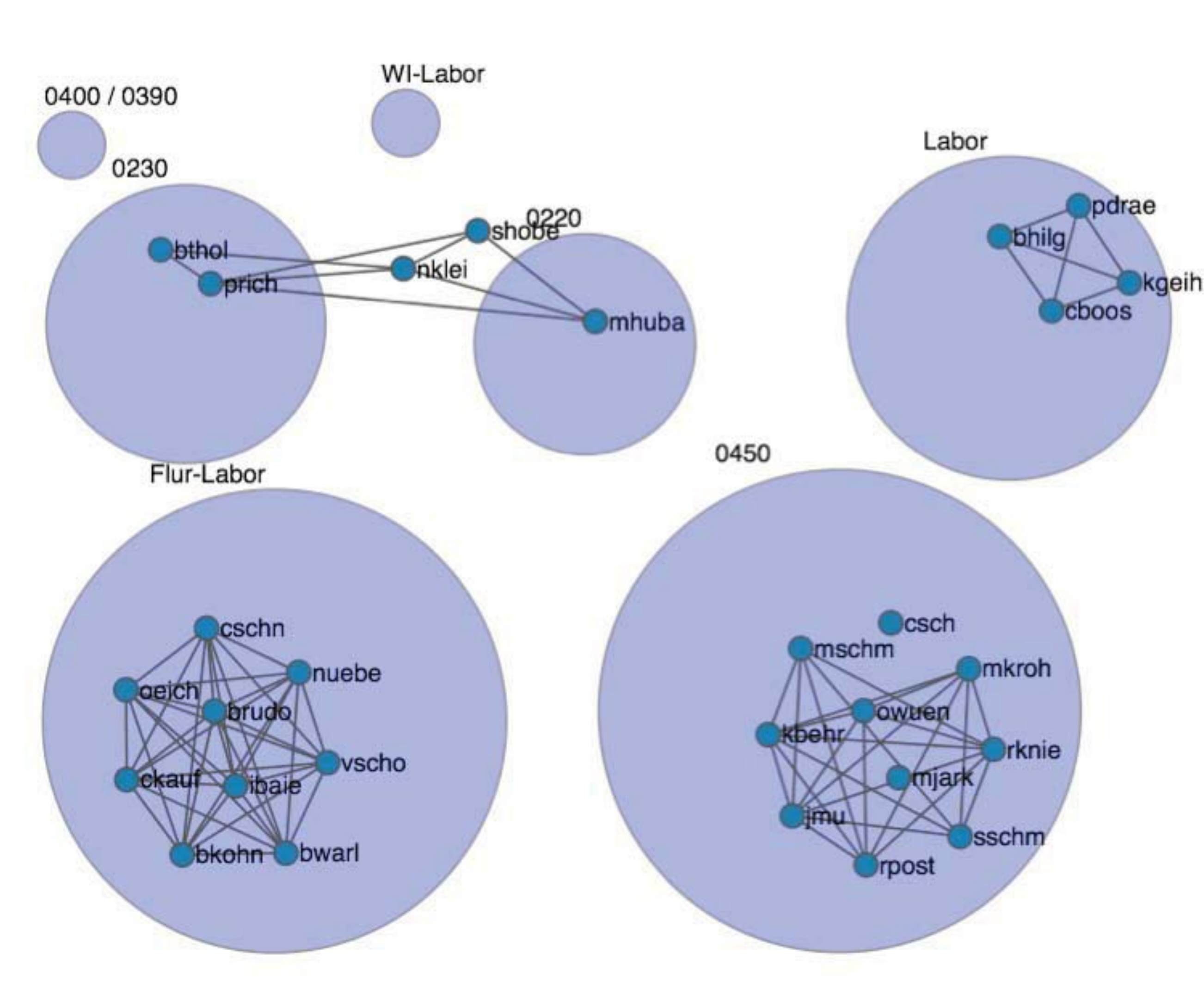}
  \caption{A screenshot~\cite{ABDHKMMMSS:12} of the \emph{map view} of \mygroup. The large circles denote individual rooms, the smaller circles participants; connections between those indicate ongoing face-to-face contacts.}
    \label{fig:locations}
\end{figure}

\subsubsection{Features of \conferator and \mygroup.}

\conferator and \mygroup enable ubiquitous social networking using profile information, data from social contacts, as well as from monitoring the current activity streams, \eg ongoing contacts, visits of talks, \bibs posts, \twitter tweets, etc.
From a data mining perspective, we can \mbox{exploit} the collected social structures provided by both social networks and social resource sharing systems for supporting complex and structured interactions: We can recommend persons, for example, based on joint research topics or contacts.
We apply data mining methods on the collected data to make this information visible to our users. Different trust and privacy settings, \eg concerning the visibility of contacts and locations, allow a selective distribution of sensitive information which is important for increasing trust in the system~\cite{SHHL:12,CEGHKLNRSSSW:12}. 

Conference participants can recall their individual contacts, \eg as virtual business cards, \cf~Figure~\ref{fig:peerRadar}. The system allows the setup of a complete profile,  social networking to other participants, and the management and personalization of the conference schedule, by providing helpful information about the individual talks, upcoming talks, and the ability to pick talks for a personal conference program. In addition, recommendations for contacting \emph{interesting} persons are provided.
\conferator and \mygroup feature the following basic options: They allow participants to recall their contacts (in the \emph{contact view}, depicting the shares of contacts to other users), to observe the (online) social live around them (\emph{timeline view}, as a configurable list of time-based events and activities), to help in finding participants via the localization component (\emph{map view}, see Figure~\ref{fig:locations}) and to browse the individual profiles (\emph{profile view}, see Figure~\ref{fig:peerRadar}).
 Furthermore, recommendations and context-specific notifications are provided~\cite{ABDHJMMSS:11,ABDHKMMMSS:12}, \eg concerning other conference participants or the contact history.
 
In addition to the profile pages, the timeline visualization~\cite{ABDHKMMMSS:12} is one of the main visualizations of \conferator. It is an aggregation of different events and activities of the participants arranged in a time-oriented list. This includes, for example, contacts, \bibs posts or \twitter tweets: It provides an aggregated view on the currently active topics published on \twitter or Bibsonomy by the participants and the conversations that recently happened. \mygroup utilizes the same visualization technique: The timeline, for example, which is displayed on a large LCD screen at the KDE group at the University of Kassel, often stimulates interesting research discussions and enables enhanced dissemination and exchange of knowledge.
The systems are continuously refined according to user feedback and usability studies, leading to continuous improvement of the systems and implementation of new useful features.

Altogether, \conferator and \mygroup create a ubiquitous and social environment where large groups (implicitly) collaborate together using electronic media to accomplish certain tasks. Ultimately, this enables a form of \emph{Collective Intelligence}~\cite{Leimeister:10,MLD:09,MLD:10,Weschsler:71}: A very intuitive notion captures the term as
``groups of individuals doing things collectively that seem
intelligent''~\cite{MLD:09}. In the context of conferences and working groups, for example, talk and contact recommendations can be improved using the collected interaction data. Furthermore, interesting topics can be identified at a conference, for example, by taking the top visited talks and their descriptions into account.

%%%%%%%%%%%%%%%%%%%%%%%%%%%%%%%%%%%%%%%%%%%%%%%%%%%%%%%%%%%%%%%%
%%% Section: Analysis
%%%%%%%%%%%%%%%%%%%%%%%%%%%%%%%%%%%%%%%%%%%%%%%%%%%%%%%%%%%%%%%%
\section{Analysis of Social Interaction Networks}\label{sec:analysis}

The analysis of online social network data has received significant attention: As a prominent example, Mislove \et~\cite{MMGDB:07}, applied methods from social network analysis as well as complex network theory and analyzed large scale crawls from prominent social networking sites. They worked out properties common to all considered social networks and contrasted these to properties of the web graph. Broder et al.~\cite{BKMRRSTW:00} utilized complex network theory for analyzing (samples from) the web graph. Kwak et al.~\cite{KLPM:10} provide an an analysis of fundamental network properties and interaction patterns in \twitter, while Ahn et al.~\cite{AHKMJ:07} provide an analysis of the topological characteristics of networks in online social networking services.
Furthermore, Newman~\cite{newman2003social} analyzed many real life networks, summing up their characteristics. Similarly, an analysis of fundamental network properties and interaction patterns in \twitter can be found in~\cite{KLPM:10}.

In this section, we investigate structural interrelations between social interaction networks, and analyze communities and roles in face-to-face contact networks at conferences. Furthermore, we inspect and investigate the interactions and dynamics of conference participants. In this way, we aim at gaining a better understanding of the behavior and its underlying processes. The extracted knowledge can then be applied, for example, for optimizing processes, or the integration into applications. 

For enhancing our understanding of social interaction networks, we focus on the network structures and the properties of the nodes. In the context of conferences, for example, we aim to analyze the behavior and relations of the participants in order to ultimately uncover common patterns and trends throughout the conferencing scenario. We combine data mining and social network analysis techniques for examining social interaction networks and summarize specific methods and analysis results in the context of those interaction networks. We present analysis results in the context of social bookmarking and social resource sharing systems, as well as in human face-to-face contact networks -- using the \conferator and \mygroup systems.

In the following, we first present an analysis of structural interrelations on social interaction networks. After that, we discuss dynamics of communities and roles in human contact networks at conferences and present an analysis of dynamic and static behavior of conference participants.

\subsection{Structural Network Interactions and Correlations}\label{sec:analysis:structural}

Social interaction networks are of large interest for major applications, such as recommending contacts in online social networks or for identifying groups of related users, \cf~\cite{MABHS:11,51seitenPaper}.
There, we provide a detailed structure and semantic based analysis of three sample social media applications: \twitter (microblogging), \flickr (social resource sharing), and \bibs (social bookmarking). In each application, we identify various implicit and explicit social interaction networks, also called evidence networks, focusing on explicit and implicit user traces.

In the following, we outline two main issues: Are there interrelations and correlations between the interaction networks? Furthermore, can these be applied for the analysis and data-driven assessment of communities?
The second question is especially important, since one of the main problems of community detection~\cite{Newman:04,Newman2006fcs,LF:09,FC:07} is the non-trivial evaluation and validation of the identified communities. The assessment of the quality of a given community is always application dependent and \emph{relative} to certain aspects of user relatedness, \eg race of individuals in \cite{Newman:03}, shared topical interests in social bookmarking systems, or social traces manifested in the social interaction networks. Often there is no gold-standard evaluation data at hand in order to validate the discovered groups.

\subsubsection{Interrelations in Social Interaction Networks.}
As a starting point for the analysis of the considered social interaction networks, we perform a structural interrelation analysis in order to identify general properties and to compare different networks focusing on common network structures, community structures, and user-relations within these networks.
We examine and compare social interaction networks applying user data from the real-world social bookmarking application \bibs, \flickr and \twitter.
We analyze general structural properties of the obtained networks and comparatively discuss major structural characteristics in order to show that there are structural and semantic inter-network correlations between the different evidence networks.

In particular, we examine several general structural properties, the degree distribution and the degree correlation, indicating significant similarities of the networks. Furthermore, we analyze topological and semantical distances, the dependencies of the networks' neighborhood, and the inter-network correlations, and collect evidences for strong correlations and interrelations between the considered social interaction networks.
Specifically, we analyze inter-network correlations between such user-generated networks and show that these relations are strong enough for inferring reciprocal conclusions between the networks.

\subsubsection{Analysis of Community Structure.}
The social interaction networks are thoroughly analyzed with respect to the contained community structure. Using standard community measures, \eg modularity~\cite{Newman:04},  segregation index~\cite{Freeman:78}, and conductance~\cite{LHK:10}, we show that there is a strong common community structure accross different social interaction networks.
Furthermore, we analyze the rankings between a large set of communities mined on the different networks, and show, that the induced rankings are reciprocally consistent.
Therefore, since the correlations and dependencies are strong enough for assessing structural analysis techniques, \eg community mining methods, implicitly acquired social interaction networks can be applied for a broad range of analysis methods instead of using expensive gold-standard information.
We therefore propose an approach for (relative) community assessment: The presented approach is based on the idea of \emph{reconstructing existing social  structures}~\cite{SS:09} for the assessment and evaluation of a given clustering.
This provides for a simple yet cost-efficient assessment option, since we can apply secondary data, \ie implicitly acquired social interaction networks capturing user relatedness, for assessing community structures for another network in the same application context.
Specifically, the presented analysis is thus not only relevant for the evaluation of community mining techniques, but also for implementing new community detection or user recommendation algorithms, among others.

\subsection{Communities and Roles in Face-to-Face Contact Networks}\label{sec:analysis:communities}

Gaining a better understanding of communities and roles in social interaction networks helps for a number of applications, for example, for personalization, community detection, or recommendations.

In ~\cite{ADHMS:12}, we investigate user-interaction and community structure according to different special interest groups during a conference. For the analysis, we considered the contact network of the LWA 2010 conference that we obtained using the \conferator system. The analysis thus utilizes real-world conference data capturing community information about participants and their face-to-face contacts, and is grounded using information about membership in special interest groups and academic status/position. We analyze various general structural properties of the contact graph, confirming previous results of the SocioPattern experiments concerning typical face-to-face contact networks at conferences \eg~\cite{DBLP:journals/corr/abs-1006-1260}.

Furthermore, we examine different explicit and implicit roles of conference participants in the context of community structure at the conference.
Role mining concerning communities mainly analyzes the relations between the communities for a specific actor. Scripps et al~\cite{STE:07a,STE:07b} present a method for assessing roles with respect to the membership in the communities and the potential to bridge or to connect different communities. In this way, different actor profiles concerning their centrality prestige and their community importance can be derived. Chou and Suzuki present a similar method considering a set of given communities~\cite{CS:10} for such a community-oriented analysis.

\subsubsection{Community Dynamics.}
For analyzing the community structure and its dynamics, we consider community detection methods in a time-based analysis using the special interest group communities as a ground truth. We analyze, whether there is a detectable community structure in the investigated face-to-face networks that is consistent with the one given through the groups.
Using the community measure \emph{segregation index}~\cite{Freeman:78} and modularity~\cite{NG:04}, for example, we can observe the trend, that more relevant (\ie longer) conversations are biased towards dialog partners with common interests, as captured by the interest group membership: Members of a special interest group tend to talk more frequently and longer within their interest group, as analyzed on the accordingly weighted social interaction networks.

Furthermore, with increasing minimal conversation lengths the induced social interaction networks show a more pronounced community structure, both considering automatically mined communities as well as the communities defined by the special interest groups. This finding is also supported by analyzing the densities in the respective subgraphs induced by the community detection method, and the respective ground-truth communities.
Therefore, the face-to-face contacts show inherent community structure, which is also consistent with the special interest groups.

\subsubsection{Roles and Communities.}
\begin{figure}
  \centering
  \includegraphics[width=1.0\columnwidth]{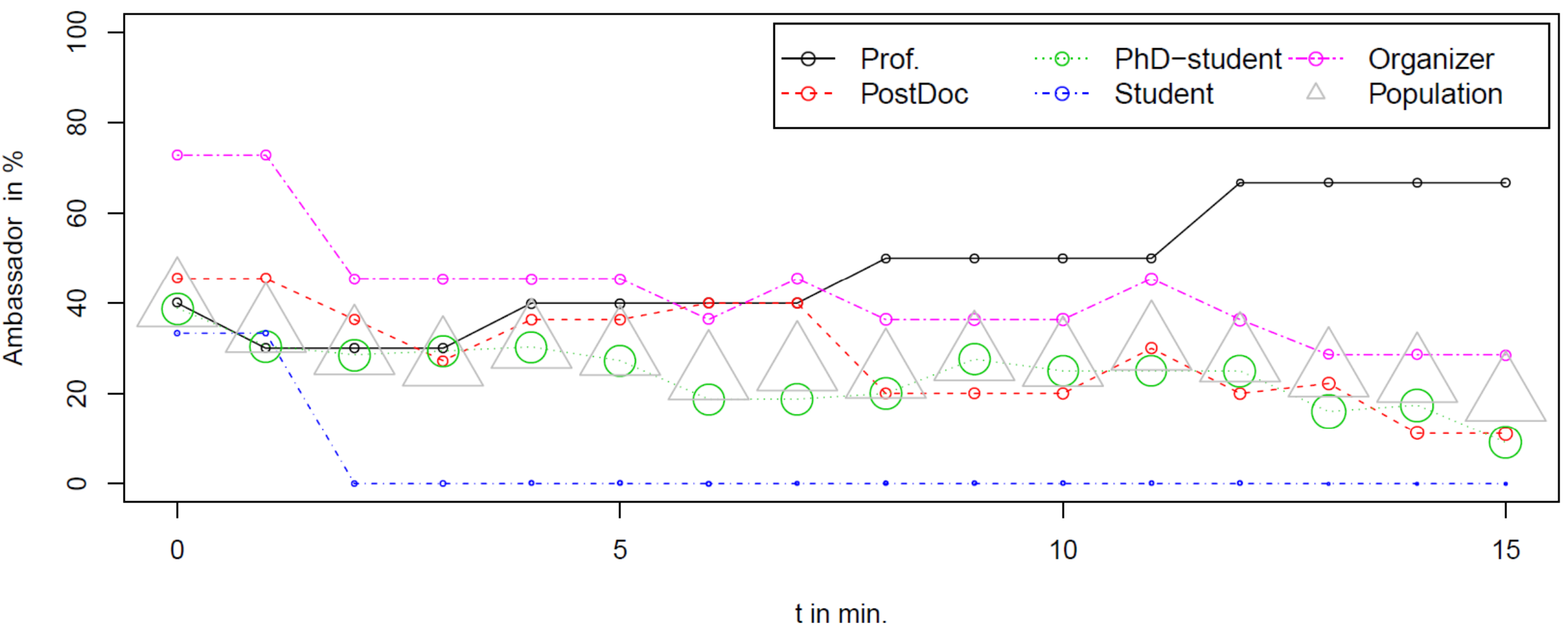}
  \caption{Fraction of the participants that assume the role Ambassador at LWA 2010 when considering all conversations with a length $\geq t$, \cf~\cite{ADHMS:12}. Ambassadors are characterized by a high degree and by a high connectivity between different communities, \cf~\cite{STE:07a}.}
\label{fig:lwa2010-ambassador-role}
\end{figure}

For analyzing the dynamics of roles of the participants, we also utilize the ground-truth community information. We provide a time-based analysis of the roles, and different role profiles. For example, we consider \emph{bridges} connecting communities, or \emph{ambassadors} as important bridging actors.
As an example, Figure~\ref{fig:lwa2010-ambassador-role} shows the \emph{Ambassador} role in a time-based analysis concering increasing minimal conversation length thresholds. Intuitively, an ambassador has a high node degree and is able to connect different communities. In Figure~\ref{fig:lwa2010-ambassador-role}, we observe the trend that the organizers start out as ambassadors for shorter conversations, but become less and less important for longer conversations. In contrast, more and more professors are categorized as ambassadors, for longer conversations. For organizers, this is in line with our expectations, since they need to be involved in a lot of shorter conversations. On the other hand, especially professors seem to be able to bridge communities with longer (more meaningful) conversations.
Similar results concerning the communities and roles were also obtained for other conferences, \eg for the Hypertext 2011 conference~\cite{MSAS:12}.  

\subsection{Structure and Dynamics on Interactions of Conference Participants}\label{sec:analysis:anatomy}

Understanding the mechanisms of conference interactions and their dynamics, \eg using a time-based analysis of conference participants, can help in many ways: It can increase the efficiency and effectiveness of individual networking, support the conference organization, be utilized for process optimization or be incorporated into advanced data mining methods and tools.
However, the analysis of conference interactions is not easy if conventional tools like questionnaires are used, \cf~\cite[p. 45 ff.]{WF:94}, since then mostly \emph{static} analyses of the behavior and processes can be performed, while the \emph{dynamic} nature of conference interactions is not accounted for.

\begin{figure}[!t]
\centering
\includegraphics[width=.9\columnwidth]{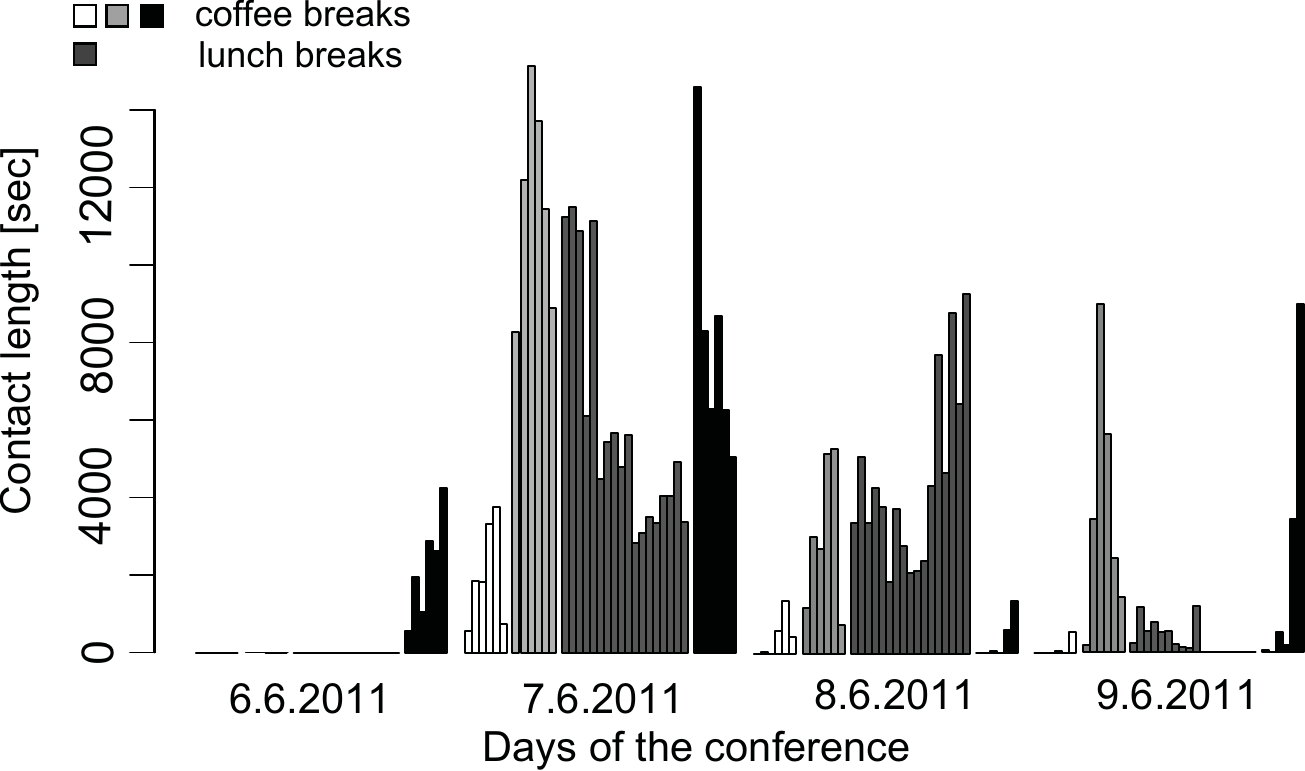}
\caption{Dynamic contact activitiy during the ACM Hypertext 2011 conference~\cite{MSAS:12} - focusing on the coffee and lunch breaks. The time slices contain contact durations for the complete conference except for the sessions. The start times of the coffee breaks were as follows: 8:30, 10:30, 15:30 and 16:00. Their duration was always 30 minutes. The start and duration of lunch breaks varied. Except for the last day all started at 12:30 and took at least one hour. Each bar represents a 5 minute window summing up all durations of all face-to-face contacts contained in this window; adjacent bars belong to the same coffee or lunch break.}
\label{fig:complconf}
\end{figure}

In~\cite{MSAS:12}, we present an in-depth analysis of the static and dynamic nature of a conference, exemplified by the ACM Hypertext conference 2011 in Eindhoven, The Netherlands, where we collected RFID face-to-face contact data using the \conferator system.
Figure~\ref{fig:complconf}, for example, depicts the conference contact dynamics for all coffee and lunch breaks, and gives a first impression of the contact hotspots during the conference. Since the setup of the \conferator system started in the middle of the first day (6.6.2011), all previous time slices are empty. 
As expected, there were a lot of interactions between participants which decreased over time as the conference progressed. This can partly be explained by departing participants who were returning their RFID tags. The short peaks at the last two coffee breaks are also an exception and might be explained by the conference attendees saying goodbye to each other.

In the following, we specifically consider the \emph{individual} behavior of participants at the conference, and their \emph{community} interactions, \eg insights into the communication in tracks.
One of the first experiments in a similar context at conferences was performed by Cattuto and colleagues, \cf~\cite{ASCBCB:09,BCSBA:10,SCBBA:10}. We extend their findings with a number of results for homophily and session attendance of the participants, their communication behavior and an analysis concerning their submitted papers.

\subsubsection{Individual Behavior.}
Our analysis focuses on individual behavior considering the communication within tracks, after the end of a session, and the behavior of special roles, \eg presenters. Focusing on these, we also include the content of the submitted papers using a bag-of-words model. In contrast to intuition, in the analysis of the presenters, we cannot confirm our assumption, that these were more involved in talks with participants presenting similar work based on the content of their papers. For the track visiting behavior, we find that all tracks focus on their own community, concerning the session attendance of the individual members of the track. Furthermore, we provide an in-depth analysis of participants, presenters, session chairs, concerning their roles at the conference, by mining role patterns, \eg for the ambassador or bridge role, \cf~Section~\ref{sec:analysis:communities} for details. The strength of the affiliation of the conference turns out to be one of the strongest features in the patterns that determines to connect different communities that are present at the conference.

\subsubsection{Community Interactions.}
\begin{figure}[htb]
  \centering
  \includegraphics[width=0.88\columnwidth]{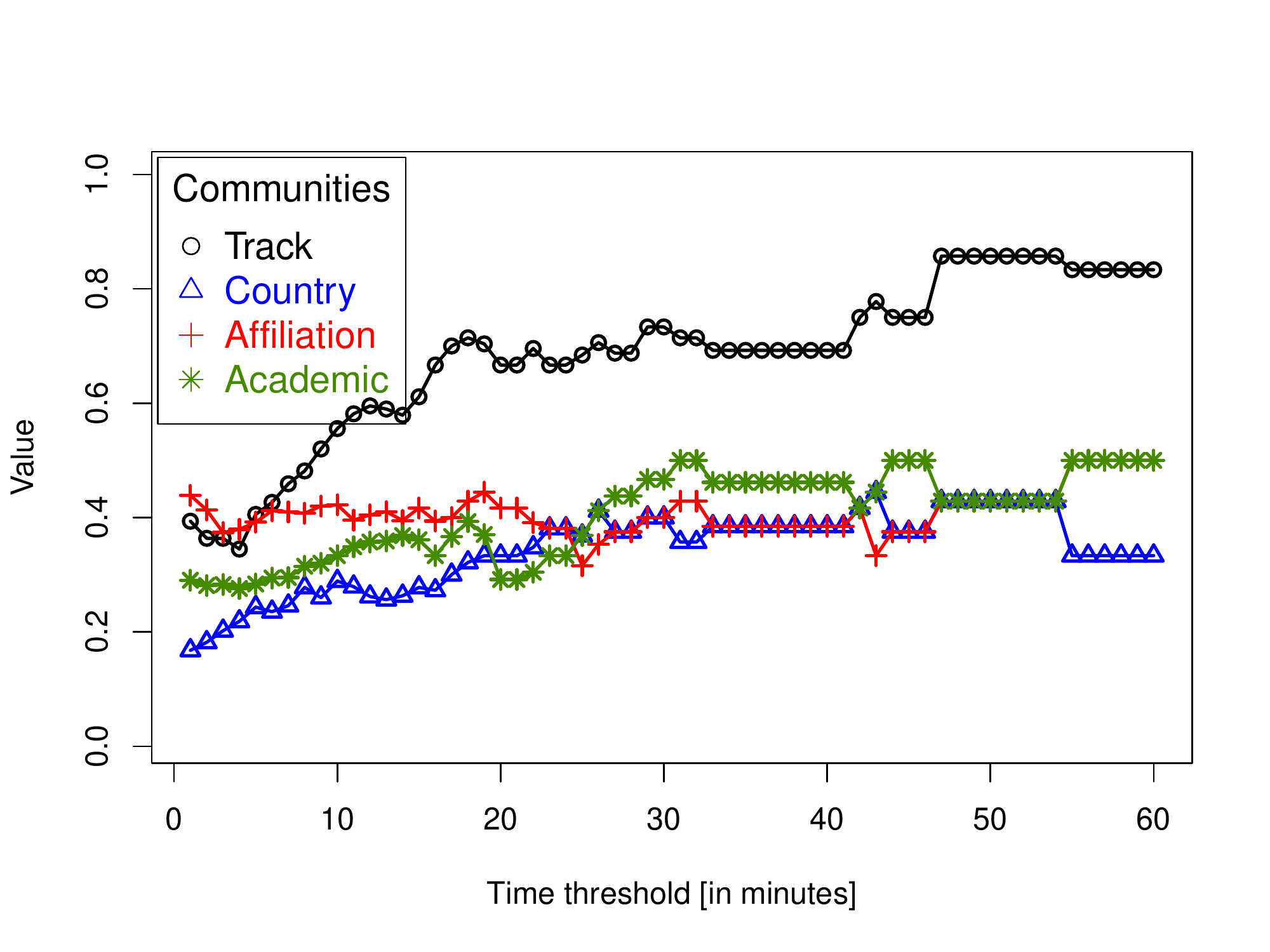}
  \caption{Overview: Community quality indicator ($p$-value~\cite{STE:07a}) for the partitionings track, country, affiliation and academic status, in a threshold-based analysis using different minimal conversation lengths, \cf~\cite{MSAS:12}. The higher the $p$-value, the higher the probability for a contact within a community.}
  \label{fig:pValue}
\end{figure}

For the analysis of interactions within different communities, we investigate different partitionings, \eg concerning the individual tracks and sessions, but also automatically mined communities with respect to their contact behavior. Figure~\ref{fig:pValue} shows an example of community structure concerning different 'organizational' aspects, \ie \emph{track}, \emph{country}, \emph{affiliation to the conference} and \emph{academic status} communities: These are partitionings according to the visited track, the country of origin, the affiliation to the conference, and the academic status, \eg Professor, PhD, or student. In Figure~\ref{fig:pValue} we observe, for example, that the length of the conversations has a high impact for the track community: The longer the conversation, the higher the probability of having a contact within the same track community.
Vice versa, we also observed, that longer conversations are more probable, if the dialog partners are both members of the same track. 

%%%%%%%%%%%%%%%%%%%%%%%%%%%%%%%%%%%%%%%%%%%%%%%%%%%%%%%%%%%%%%%%
%%% Section: Predictive Modeling
%%%%%%%%%%%%%%%%%%%%%%%%%%%%%%%%%%%%%%%%%%%%%%%%%%%%%%%%%%%%%%%%
\section{Predictive Modeling }\label{sec:predictive}

Approaches for predictive modeling aim to learn models for later deployment, \eg for estimating a function that predicts a certain value for future cases~\cite{FPS:96a,FPS:96b}.
Classification or regression are such typical predictive tasks. In this section, we present predictive approaches in the context of social interaction networks utilizing data from social and ubiquitous systems, specifically the \conferator and \mygroup systems. For these, standard predictive methods, \eg classification using random forests, as well as \pagerank-based and link mining approaches were adapted and extended as needed.

Below, we present approaches for resource-aware localization, for recommendations of experts, and for analyzing the predictability of links in face-to-face contact networks.
For the localization, we combined tracking signals and contact information from social interaction networks for enhancing the performance of common machine learning techniques using different voting mechanisms. Further, we adapted the well-known \pagerank method~\cite{BP:98} on special social interaction networks. These are mediated by software (source code) resources using information from revision control systems, extended with social contact information. Using this information, the performance of the recommendation method is significantly improved compared to standard approaches.
Finally, we consider a link prediction approach in social interaction networks where we analyze the impact of stronger ties and identify influence factors for the prediction.

\subsection{Resource-Aware RFID Indoor Localization utilizing Social Interactions}\label{sec:predictive:localization}

In the context of \conferator and \mygroup, knowing where attendees and colleagues are, respectively, supports group organization and thus facilitates everyday work processes.  The localization component provides the locations of all users, and shows where their conversations take place.
During conferences, for instance, \conferator offers the possibility of observing who is visiting a given talk, thus facilitating the academic exchange during the subsequent coffee breaks.

Furthermore, it is possible to identify hotspots, \eg conference rooms where a large number of conference participants is listening to -- apparently interesting -- talks, and to potentially recommend those to undecided participants.
Capturing and visualizing live interactions of individual users is an important task for \conferator and \mygroup, essentially in order to enable collective intelligence. Therefore, a localization framework is a central component for such a system.

\subsubsection{Resource-Aware Localization Setting.}
In~\cite{SDAHS:11}, we present an approach for a resource-aware and cost-effective indoor localization method in the context of RFID based systems.
While approaches for outdoor localization can utilize various existing sources, \eg GPS signals, mobile broadcasting signals, or wireless network signatures, methods for indoor localization usually require special installations (\eg RFID or Bluetooth readers), and/or require extensive training and calibration efforts.

The proposed cost-effective and resource-aware solution requires only a small number of RFID readers. Furthermore, our method can be applied to installations, where readers cannot be positioned freely. 
The latter constraint is encountered often, especially in historical buildings under monumental protection.

In contrast to typical experiments that examine RFID localization in laboratory experiments, \eg~\cite{NLLP:04,HVBW:00}, we present an analysis of data collected in a real-life context, that poses additional problems in terms of signal quality and noise: We consider a real-life localization problem at room-level, \ie the task to determine the room, that a person is in at a given point in time.

\begin{figure}
  \begin{center}
    \includegraphics[width=0.66\columnwidth]{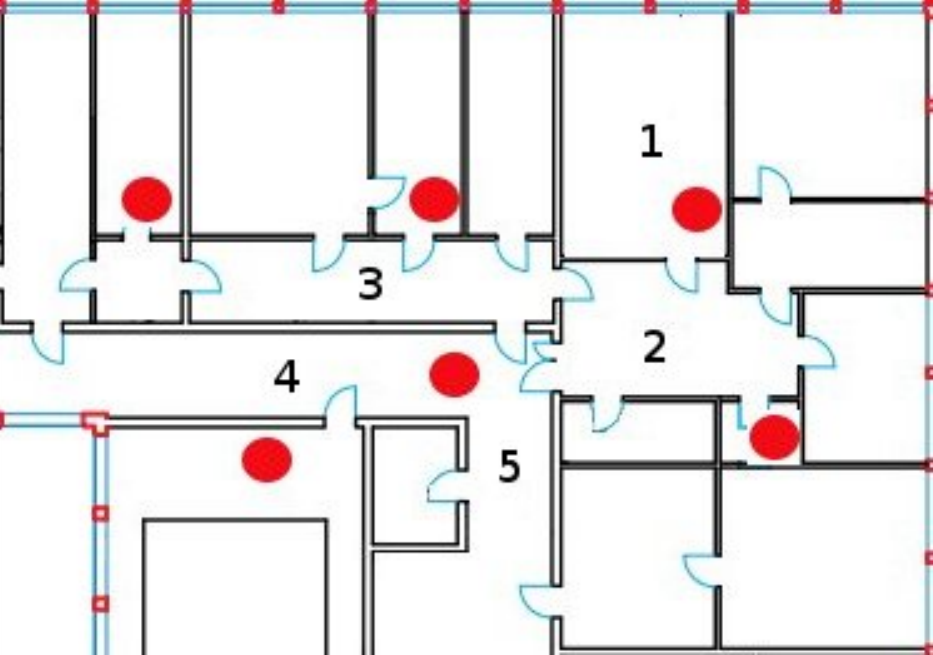}
    \caption{Example conference area~\cite{SDAHS:11}: The numbered rooms were used by participants during the poster-session of LWA 2010, the circles mark the positions of RFID readers.}
    \label{lwa2010-figure_conferencemap}
  \end{center}
\end{figure}

\subsubsection{Social Boosting: Improving Localization using Face-to-Face Contacts.}
We present an analysis of the contact and proximity data in order to prove the validity and applicability for the sketched application. Additionally, we evaluate the benefits of several state-of-the-art machine learning techniques for predicting the locations of participants at the room-level. We propose to  utilize the (proximity) contacts of participants for improving the predictions of a given core localization algorithm using different voting mechanisms. We evaluate the impact of different strategies considering the top performing machine learning algorithm. The real-world evaluation data was collected at LWA 2010, \cf Section~\ref{sec:analysis:communities}. Figure \ref{lwa2010-figure_conferencemap} shows the covered conference area with $6$ readers put in adequate positions.

We evaluated several state-of-the-art machine-learning algorithms in this context, complemented by novel techniques for improving these using the RFID interaction contacts.
 The results of the experiments yielded several reasonable values for the applicable parameters. For the simpler algorithms, they could also have been learned in a short preceding training phase, which demonstrates the broad applicability of the approach in the sketched resource-aware setting.
Overall, using the social face-to-face contact information in the \emph{Social Boosting} algorithm~\cite{SDAHS:11}, we could improve the localization accuracy from 84\,\% using a baseline algorithm to nearly 90\,\%, as evaluated during the poster session of the LWA 2010 conference.

\subsection{Combining Interaction Networks for Expert Recommendation}\label{sec:predictive:recommendation}

Recommendations play an important role in supporting software development, especially in larger teams: Locating experts for a given problem is one of the main challenges when working in a large team.

\subsubsection{Expert Profiling.}
Our work~\cite{MAS:11} presents an approach for expert profiling and recommendation in such contexts:  First, specific profiles of developers concerning resources, packages, and projects provide an overview on the area that the respective developer is working on, \eg for an overview on the activity of a team. Second, predictive models, \eg modeling the familiarity of developers with specific resources, can increase the effectiveness of other team members. In the case of specific questions,  persons can be suggested that are especially familiar with these resources. In this way, knowledge management and knowledge transfer, \eg transfer of projects, instructing new team members, or participation in open-source projects, can be successfully implemented.

In the context of \mygroup, we focussed on supporting software development groups~\cite{MAS:11}. The presented approach can potentially be generalized for any organization using revision control systems, \eg for recommending collaborators based on changes in documents, papers, or wikis, etc.
We propose an approach for analyzing the communication and commit structure of a development group. In our case study, the development group uses CVS as a code versioning system; additionally, conversations between developers are captured using the RFID tags applied by \mygroup as described above.
In the sketched scenario, the two basic assumptions are the following: The number of added and removed lines of code  that a developer commits for a specific resources serves as a proxy for her \emph{familiarity} with this specific portion of code, \eg~\cite{IWPSE2005,Hindle2008}. Additionally, conversations between developers serve as a way for transferring \emph{knowledge} from one developer to another. Therefore, such interactions also help to increase the familiarity of developers with the source code.

\subsubsection{Combining Interactions for Recommendation.}
\begin{figure}[htb]
   \centering
   \subfigure[Exemplary resource tree with edges from seven files to three developers~\cite{MAS:11}. The edge weights are determined by the hierarchical partitioning of the lines of code in each resource, and by the contribution of each developer (at the bottom of the tree), respectively.]
   {\includegraphics[width=0.48\linewidth]{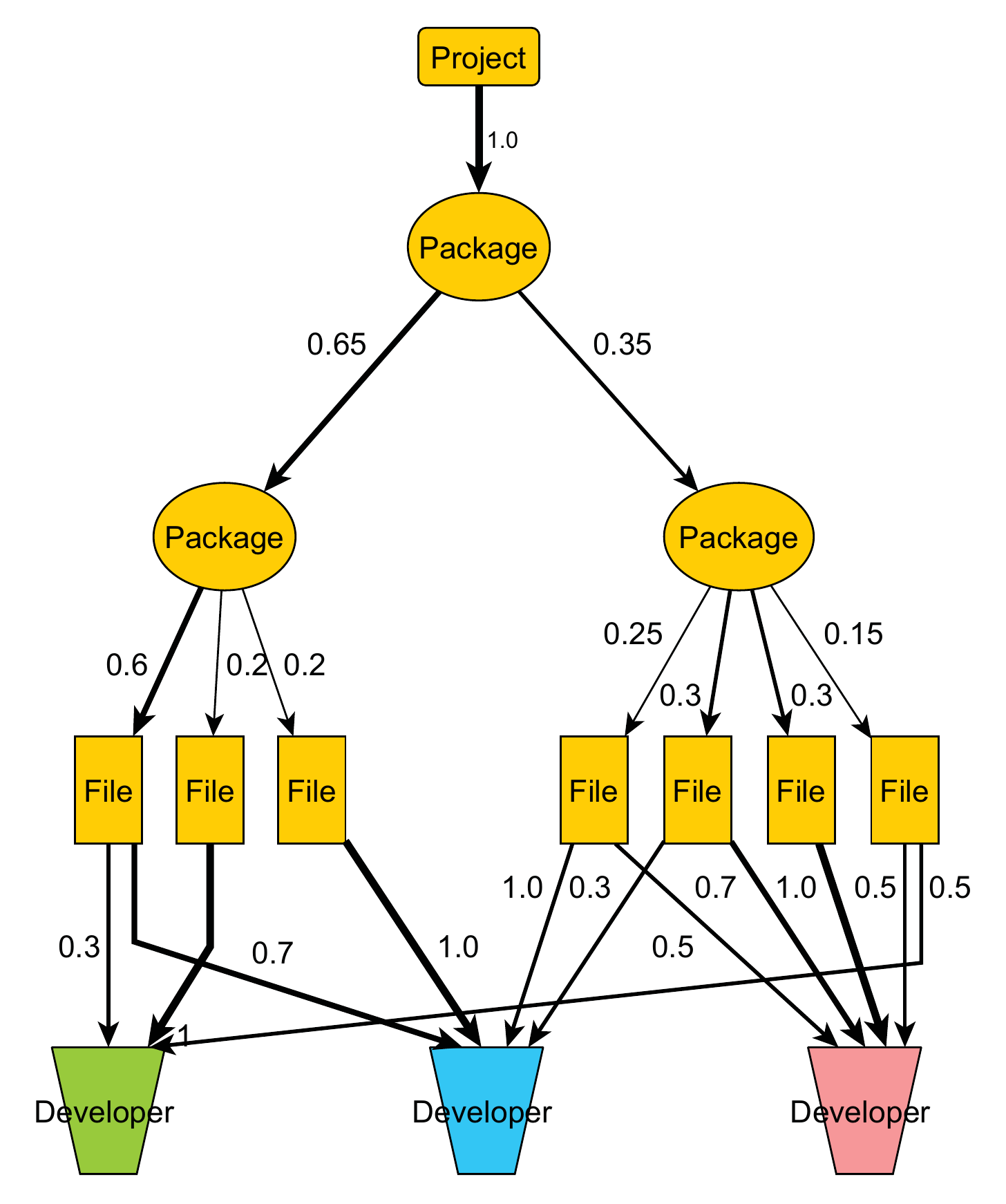}}
   \hfill
   \subfigure[Example developer contact graph~\cite{MAS:11} (the dashed lines correspond to those in the resource tree). Edges between developers are weighted by their relative conversation durations, normalized by a parameter (0.1 in our example) which aims to model the amount of knowledge transferred in communication.]
   {\includegraphics[width=0.48\linewidth]{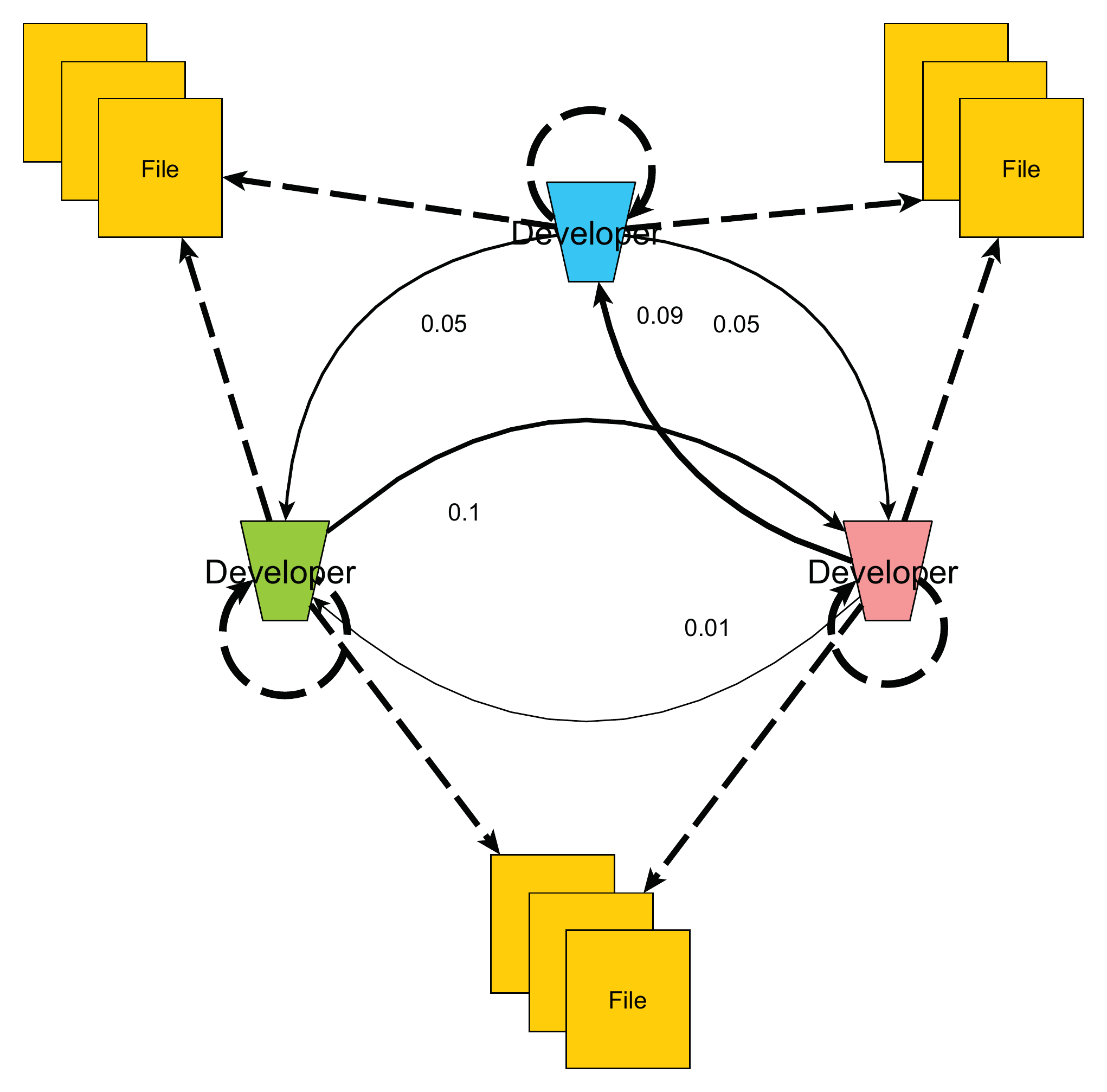}}
   \caption{Exemplary resource and developer contact graphs~\cite{MAS:11}. For recommending software developers, the PageRank algorithm~\cite{BP:98} is then applied to their combination.
   }\label{fig:recommenderStructures}
\end{figure}

We provide two novel classes of graphs built from structural data for mining developer and resource profiles concerning the 'familiarity' with specific resources, packages and/or projects.
We analyze code changes and the structure of the software projects. In this way, we create resource trees resembling the hierarchic organization of source files, \cf Figure~\ref{fig:recommenderStructures} for a simple example. The contribution of each developer is then measured by the number of changed lines of code.
We combine this information with the RFID face-to-face contact network measuring the face-to-face contacts of the developers for modeling the real-life communication, especially conversations that take place before a commit. Using this social interaction network, we are able to store the time and duration of a conversation, so that this information can be analyzed further. This enables the capture of conceptual knowledge which is mostly propagated via conversations and cannot be extracted by mining software repositories. In addition to weighted edges which reflect the relative amount of changed lines of code, we consider edges between the developers. These edges are weighted by the cumulative duration of the face-to-face contacts of the developers within the last eight hours before committing changes to the source code. Thereby, we connect their real-life communication and social interactions using the \mygroup system. The resulting structure captures important knowledge of a social group: Exogenous information, \eg developers writing code by themselves, and endogenous information representing the knowledge transfer from one individual to another by means of communication.

Utilizing the graph structure, the \emph{PageRank}~\cite{BP:98} algorithm is then applied: We can calculate either a set of developers that are familiar with a specific portion of source code, or we can analyse the quality of code coverage by a given subgroup of software developers.
We evaluated the approach in the context of a developer group with a medium sized project comparing the proposed approach against a ground-truth expert-ranking. 
As shown in the evaluation, the explanatory and predictive power of exogenous experience captured by CVS logs combined with the endogenous flow of experience captured by logged communication shows strong improvements concerning an approach only based on a lines-of-code analysis.
The evaluation results demonstrate the effectiveness and impact of the proposed approach: The contact information (RFID) usually improves the performance of the proposed approach compared to a baseline using only a lines-of-code analysis.

\subsection{Link Predictability in Face-to-Face Contact Networks}\label{sec:predictive:contacts}

Link prediction~\cite{LK:03} in social network considers the dynamics and mechanisms in the creation of links between the actors of social networks. The goal is to learn a model for predicting new and/or recurring links accurately. This also relates to mobility~\cite{WPSGB:11,Barabasi:02} and dynamic behavior~\cite{WS:98,TB:09}. Link prediction in social interaction networks has a number of prominent applications, including the prediction of missing links, \cf~\cite{LK:03}, for improving collaborative filtering, \eg~\cite{ZXC:05}, or for recommending new contacts, \eg~\cite{XC:09,PSM:11}. A method for recommending interesting contacts, for example, has been deployed in the \conferator system.

There is already a large body of research for link prediction concerning \emph{online} social networks, \eg~\cite{AA:03,LK:03,MM:07,LZ:09,Katz:53,ZLZ:09}. However, important aspects of face-to-face contact networks, \ie interactions that happen offline, still remain largely unexplored.
Sociological experiments and approaches, \eg~\cite{Lim:12,Gatica:09,EE:99}, mainly rely on questionaires, diaries, or recordings, and usually only consider rather small groups, \cf~\cite[p. 45 ff.]{WF:94}. 
In contrast, the \conferator system is able to collect face-to-face contacts (and the according networks) at much better precision and for rather larger groups.
 The analysis of such networks can potentially provide more direct answers to fundamental questions, \eg how do personal links get established, what are influence factors in this contexts, what is the impact of stronger ties in face-to-face contact networks.

\subsubsection{Predicting New and Recurring Links.}
In~\cite{SAS:12}, we aim at providing first insights for answering such questions. We focus on face-to-face contact networks. For the analysis, we apply real-world data collected at the LWA 2010 and Hypertext 2011 using the \conferator system, \cf Section~\ref{sec:analysis:communities}-\ref{sec:analysis:anatomy}.
Using this data, we can observe and analyze social interaction at a very detailed level, including the specific event sequences and durations.
We aim to predict \emph{new} contacts based on network properties of face-to-face contact networks, \eg for a recommendation setting, as an adaptation of methods for online social networks. For that, we apply and extend basic link prediction measures utilizing the network structure, and their weighted variants.

In addition, we extend the analysis in two important directions: First, we consider the length of the contacts in more detail, and analyze the impact of longer conversations. Second, we consider the prediction of future \emph{recurring} contacts, \ie renewed contacts between specific actors, \eg on the first day of the conference vs. the subsequent days. For these, we analyze influence factors and patterns for establishing such contacts, and also consider their specific \emph{durations} in a fine-grained dynamic analysis. Essentially, this leads to the analysis of the impact of \emph{stronger ties} for new and recurring contacts. We estimate the influence of stronger ties for the prediction and show its impact using real-world data of two conferences.

\begin{figure}[!t]
      
    \centering
      \includegraphics[width=1.0\columnwidth]{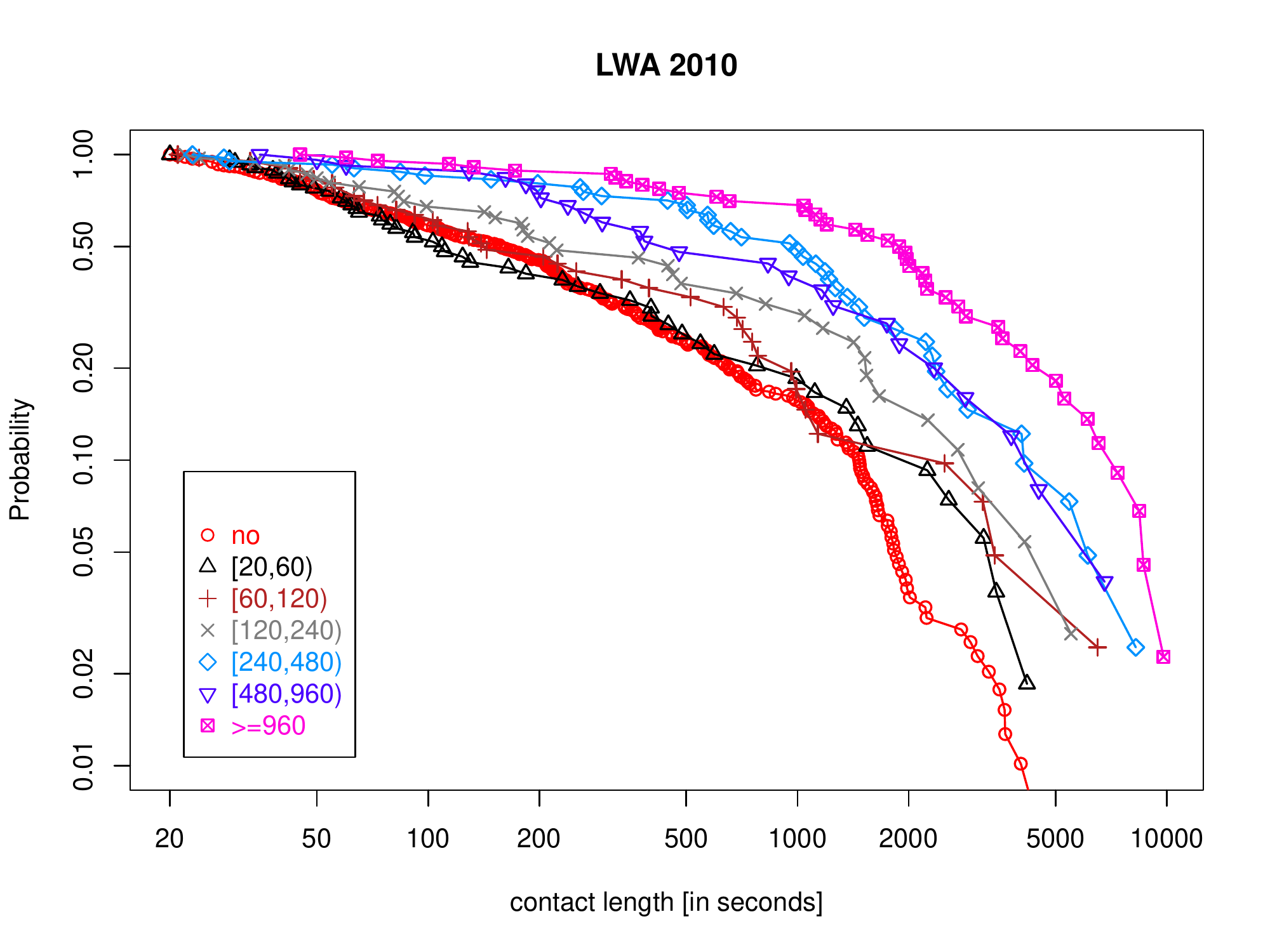}
 
  \caption{Example of predicting recurring contacts at LWA 2010 showing the impact of the contact duration between two participants on the first day, and the 
 contact length of a recurring contact at the second and third day, \cf~\cite{SAS:12}. The red line labeled with 'no' (circle symbol) in the LWA 2010 plot, for example, shows the distribution of all contacts between participants at days two and three, which had no contact at the first day. The line labeled with $[60,120)$ (cross symbol) shows the distribution of all contacts between participants at days two and three, which had a contact with contact duration between 60 and 120 seconds at the first day of the conference.} 
  \label{fig:lwa2010-impactContactLength}
\end{figure}

\subsubsection{Link Predictability and Stronger Ties.}
The results of the analysis indicate that stronger ties have a strong influence on the contact behavior and the prediction performance. As depicted in Figure~\ref{fig:lwa2010-impactContactLength}, we observe, for example, that a longer contact, \ie a conversation, is more likely, the longer the contact on the first day of the conference. An interesting further question is to find typical features to predict renewed contacts and their lengths. We show, that there are clear influence patterns of the contact durations, depending on roles such as academic status, the strength of the link to the conferences, and affiliation with the respective conference tracks. Furthermore, considering the contact durations in the ranking of the predicted contacts significantly improves the performance.

Overall, the results of the analysis provides interesting insights especially concerning the impact of the contact durations and the strength of such stronger ties.  These insights are a first step onto predictability applications for human contact networks. The indicators, patterns, and influence factors can then be integrated into more advanced prediction models in the context of face-to-face contact networks: New features can then be constructed for supervised or unsupervised link prediction methods.

%%%%%%%%%%%%%%%%%%%%%%%%%%%%%%%%%%%%%%%%%%%%%%%%%%%%%%%%%%%%%%%%
%%% Section: Descriptive Pattern Mining
%%%%%%%%%%%%%%%%%%%%%%%%%%%%%%%%%%%%%%%%%%%%%%%%%%%%%%%%%%%%%%%%
\section{Descriptive Pattern Mining}\label{sec:descriptive}

Descriptive data mining aims to uncover certain patterns for characterization and description of the data. Typically, the goal of the methods is not only an actionable model, but also a human interpretable set of patterns~\cite{Mannila:00,WJT:06}.
Descriptive pattern mining especially supports the goal of explanation-aware data mining~\cite{AR:10a}, due to its more interpretable results, \eg for characterizing a set of data~\cite{ABHS:11}, for concept description~\cite{ALKH:09b}, and for providing regularities and associations between elements in general.

In the following, we present novel methods for descriptive data mining for uncovering and extracting relations and patterns. These are applied for hypothesis generation and exploration by the user, in order to provide characteristic information about the data and networks. We focus on the mining of patterns that describe interesting communities and subsets of nodes contained in social interaction networks.

We first describe an efficient method for descriptive community mining based on the novel \CMD algorithm~\cite{AM:11a}: Descriptive community mining aims to identify interesting communities according to a community evaluation function using the network structure, and a set of descriptive attributes. Next, we discuss a generalized setting for identifying descriptive patterns utilizing exceptional model mining~\cite{LFK:08,DKFL:10,KFL:11}, and provide the \GPG algorithm~\cite{LBA:12} for fast exhaustive exceptional model mining: It can then be applied both for characterizing the network structure using an approach similar to descriptive community mining, but also for describing interesting subgroups of nodes in the network based on their properties. After that, an exploratory approach~\cite{AL:13} on social interaction networks using geo-tagged social media is presented, which utilizes both techniques presented above.
All approaches can be applied for supporting the user, \eg for recommendations, facetted browsing, or for interactive exploration.

\subsection{Efficient Descriptive Community Mining}\label{sec:descriptive:community}

Community mining is a prominent approach for identifying densely connected subgroups of the nodes contained in a network.
A community is intuitively defined as a set of nodes that has more and/or better links between its members compared to the rest of the network.
Classic community mining, \eg~\cite{GN:02,NG:04,LF:09,LLM:10}, works on a set of nodes connected by links.
In general, usually not only the density within the community is assessed but the connection density of the community is compared to the density of the rest of the network~\cite{Newman:04}, \eg using the modularity~\cite{GN:02,Newman:04,Newman2006fcs}, the segregation index~\cite{Freeman:78} or the conductance~\cite{LLDM:08} as an evaluation function. Then, cuts between communities are established in such a way as to maximize the community evaluation function
The core idea of the evaluation function is to apply an objective evaluation criterion, for example, for the modularity the number of connections within the community compared to the statistically ``expected'' number based on all available connections in the network, and to prefer those communities that optimize the evaluation function.
Usually, an optimization approach is taken that partitions the whole graph subsequently into a number of parts -- each of them is then considered as a community.
The discovered communities can then be applied, for example, for recommendations or for personalization of intelligent systems.

\subsubsection{Descriptive Community Mining.}
In~\cite{AM:11a}, we present an approach for mining descriptive community patterns according to standard community evaluation measures: The proposed method collects patterns that describe communities by combinations of features, \eg tags or topics for social bookmarking systems. We can consider, for example, groups of users interested in the topics \emph{web mining}, \emph{computer} and \emph{java}.

In this way, we aim to \emph{identify and describe} interesting communities, in contrast to standard community mining approaches, \eg~\cite{NG:04,LF:09,LLM:10}, that only identify communities as subsets of users.
In contrast to such global approaches, we focus on the discovery of local community patterns. According to the idea of local pattern mining, we do not try to find a complete (global) partitioning of the network. Instead, we consider local patterns describing local communities, so-called ``nuggets'' in the data, \cf~\cite{Kloesgen:96}. The patterns should be as exceptional as possible with respect to a given community quality measure. The pattern formalization provides an intuitive description of the community, i.e., a characterization in terms of their descriptive features. This is usually not achieved by classical community mining methods that consider the nodes of a network (\eg users in a social network) as mere strings or ids -- and provide no easily interpretable description.

\subsubsection{\CMD Algorithm.}
Our proposed approach combines local pattern mining using exceptional model mining, see Section~\ref{sec:descriptive:pattern}, and community detection: We present an efficient algorithm for mining the top-$k$ community patterns with respect to a number of standard community evaluation functions, \ie the novel \CMD algorithm: Using \emph{extended frequent pattern trees}~\cite{AM:11a}, \CMD conducts an exhaustive search by traversing a representation of the pattern search space compiled into a \emph{community pattern tree} (CP-tree).
The CP-tree is a compact version of the dataset, that also contains relevant information about the graph structure. Using this tree, the patterns can be efficiently computed using only the information contained in the tree. For pruning, \CMD utilizes optimistic estimates of the community quality functions, in a branch-and-bound fashion. Therefore, we propose suitable optimistic estimates~\cite{GRW:08,Wrobel:97} which are efficient to compute.

Our approach also tackles typical problems that are not addressed by standard approaches for community detection such as pathological cases like small community sizes. Furthermore, we focus on interpretable patterns that can easily be incorporated in a practical application, for example, for recommendations.
Since in practice the entities in a network tend to belong to a number of different communities, the presented method captures overlapping community allocations.

We demonstrate our approach on networks from \bibs.
The presented approach is not limited to social bookmarking systems and can be applied to any kind of graph-structured data for which additional descriptive features are available, \eg certain activity in telephone networks or interactions in face-to-face contacts that also utilize tags or topic descriptions for the contained relations. 
The applied optimistic estimates allow a reduction of the search space by orders of magnitude, especially using the modularity quality function.
Overall the proposed optimistic estimates show huge pruning potential for many applications, especially considering the local modularity measure as an effective tool for fast descriptive community mining.

\subsection{Fast Exhaustive Exceptional Model Mining}\label{sec:descriptive:pattern}

In the context of descriptive pattern mining, the concept of \emph{exceptional model mining} has recently been introduced~\cite{DKFL:10,LFK:08,LK:12}. It can be considered as a generalization of typical descriptive approaches like association rule mining~\cite{AS:94}, subgroup discovery~\cite{Kloesgen:96} or frequent pattern mining~\cite{Goethals:03}, and enables more complex target properties, \cf~\cite{LK:12}.
Exceptional model mining tries to identify interesting patterns with respect to a local model derived from \emph{a set} of attributes, \eg a correlation or a linear regression model. The interestingness can be flexibly defined, \eg by a significant deviation from a model that is derived from the total population or the respective complement set of instances within the population. There exist heuristic algorithms~\cite{van2010maximal} for exceptional model mining; however, these cannot guarantee any optimality of its results, in contrast to exhaustive methods. Then, efficient exhaustive methods are required due to the size of the large (exponential) pattern search space.  
Possible applications include the identification of characteristic patterns~\cite{AL:09a,ABHS:11,BAESSG:12}, analysis of node information in social interaction networks~\cite{MSAS:12,SAS:12,ADHMS:12}, or descriptive community mining approaches~\cite{AM:11a}.

\subsubsection{\GPG Algorithm.}
In~\cite{LBA:12}, we present the novel \emph{GP-growth} algorithm that can be used for mining patterns with exceptional target models \emph{exhaustively}.
We propose the concept of valuation bases allowing us to derive a new algorithm capable of performing efficient exhaustive search for many different classes of exceptional models.
We extend the well-known FP-tree data structure~\cite{HPY:00} by replacing the frequency information stored in each node of the tree by the more general concept of valuation bases: These are dependent on a specific \emph{model class} and allow for an efficient computation of the target model parameters. Intuitively, in the generalized tree (\emph{GP-tree}), each node stores a valuation basis that locally provides all information necessary to compute the target model parameters. 
We characterize the scope of the presented approach, describe properties of possible target models, and discuss its instantiations for model classes presented in literature: The applicability of the proposed approach is discussed by drawing an analogy to data stream mining, providing a constructive proof for the adaptation of the valuation base approach to other possible model classes.

\subsubsection{Evaluation.}
An evaluation of the presented approach utilizes publicly available UCI data~\cite{UCI}, as well as a social data from \flickr. Our runtime experiments show improvements of more than an order of magnitude in comparison to a naive exhaustive depth-first search. This enables the application in large social network datasets. As an example, we used a dataset obtained from flickr containing about 1.1 million instances and about 1200 tags that were used as describing attributes. Then, we aimed at identifying combinations of tags (as descriptions), for which their correlation is especially strong.
As a result, even for a search depth of 2, a simple benchmark algorithm, \ie depth-first-search, did not finish the task within two full days. In contrast, the same task  performed by GP-growth finished in about 8 minutes, using a standard office PC with a 2.2 GHz CPU and 2 GB RAM. Furthermore, even for an increased search depth of 3, the task could be completed within 10 minutes. Thus, \GPG provides for a fast efficient method for exceptional model mining. Further speedups can then be achieved using optimistic estimates, \eg as discussed above in Section~\ref{sec:descriptive:community}.

\subsection{Exploratory Pattern Mining on Social Media}\label{sec:descriptive:exploratory}

Since descriptive pattern mining characterizes and summarizes the data, it can also be applied for hypothesis generation in order to derive semi-automatic and interactive exploratory approaches.

\subsubsection{Exploratory Pattern Mining.}

In~\cite{AL:13}, we present such an interactive exploratory approach on social interaction networks incorporating location and tagging information. In this context, location information is considered as a proxy for social relatedness and interaction, \cf~\cite{MASH:13,scellato2011sociospatial,mcgee2011geographic,kaltenbrunner2012eyes,AHKMJ:07} for more details. Furthermore, using tagging information assigned to the nodes of the network, we can also explore tag-similarity measures, \eg~\cite{cattuto2008semantic}, as a proxy for the relatedness and interaction. 

We present a two way perspective on exploring locations, tags, resources and their induced interaction networks: First, we aim to describe sets of related resources (\eg photos) using location-information and tags, which are semantically related as well as focused on certain locations. We can imagine, for example, browsing the map of Germany and taking an overview on the general Berlin/Brandenburg area in terms of tag descriptions. Second, we characterize given locations using tagging patterns and photos for interactive browsing. A user may click on a map to specify his point of interest, for example, and is then provided a set of tags that are used for that region.

\subsubsection{Implementation.}
We propose an iterative two step approach for the exploration of locations and resources:
The first step uses pattern mining techniques, \eg~\cite{AM:11a,AL:09a} to automatically generate a candidate set of potentially interesting descriptive tags. 
For a flexible characterization of locations at different levels the search can be adapted by employing different location-based target measures for pattern mining. 
The result of the first step is thus a set of descriptive \emph{interesting} patterns.
In the second step, a human explores this candidate set of patterns and introspects interesting patterns manually by browsing and viewing various visualizations.
The pattern mining parameters can be adapted in an exploratory fashion.
In this way, we obtain an overview on the resources in terms of their location and describing tags. Furthermore, we can characterize different regions, areas or specific locations in terms of such descriptive information.
The resulting patterns can be exploited by providing different visualizations and browsing options. Additionally, they can be filtered according to different interestingness criteria defined by the applied quality function. Furthermore, background knowledge, \eg on semantically equivalent tags, can be manually refined and included in the process.

We demonstrate the impact and validity of the presented approach in a case study using publicly available data from the social photo sharing application \emph{Flickr}.
We apply \CMD for descriptive community mining (see Section~\ref{sec:descriptive:community}) and the efficient method SD-Map*~\cite{AL:09a} for descriptive pattern mining. This could also be implemented using the exceptional pattern mining approach described in Section~\ref{sec:descriptive:pattern}.

In the case study, we show on a structural level, that the proposed approach allows us to obtain more significant communities compared to standard community detection methods as a baseline. Furthermore, the interactive approach provides a good starting point for exploratory analysis as shown by an exemplary case study.

%%%%%%%%%%%%%%%%%%%%%%%%%%%%%%%%%%%%%%%%%%%%%%%%%%%%%%%%%%%%%%%%
%%% Section: Conclusions
%%%%%%%%%%%%%%%%%%%%%%%%%%%%%%%%%%%%%%%%%%%%%%%%%%%%%%%%%%%%%%%%
\section{Conclusions and Outlook}\label{sec:conclusions}

Data mining on social interaction networks is a rather novel research area, especially considering human face-to-face contact networks. In this \work, we presented several analyses and results, examining the interaction networks in order to improve our understanding of the data, the modeled behavior, and its underlying processes. Furthermore, we showed how to adapt, extend and apply known predictive data mining algorithms on social interaction networks. Additionally, we presented novel methods for descriptive data mining for uncovering and extracting relations and patterns for hypothesis generation and exploration by the user, in order to provide characteristic information about the data and networks.
We introduced the \conferator and \mygroup applications for enhancing social interactions: Both systems have been applied in various conferences and workgroup events.

The presented data mining approaches on social interaction networks tackle several emerging research directions: First, with the increase of data in ubiquitous and social environments, the analysis and mining of this data becomes more important -- and social interaction networks provide for a valuable modeling tool in that area. Additionally, with the increasing data volume in different pervasive services and applications, the handling of big data, \ie large volumes of data, requires efficient algorithms. Furthermore, sensors are transcending into private and personal domains of life. The derivation and construction of social (interaction) networks based on the sensor measurements requires both systematic analysis and efficient and effective algorithms. The concept of \emph{reality mining}~\cite{EP:06,Mitchell:09} is then a related research direction, as a general extension of several of the presented techniques and methods. Overall, these also open up opportunities towards the \emph{ubiquitous web}~\cite{HS:10,Sheth:10}, and \emph{collective intelligence}~\cite{MLD:09,Kapetanios:08}.

%%%%%%%%%%%%%%%%%%%%%%%%%%%%%%%%%%%%%%%%%%%%%%%%%%%%%%%%%%%%%%%%%%%%%
%%%%%%%%%%%%%%%% Bibliography
%%%%%%%%%%%%%%%%%%%%%%%%%%%%%%%%%%%%%%%%%%%%%%%%%%%%%%%%%%%%%%%%%%%%%

%%%%%%%%%%%%%%%%%%%%%%%%%%%%%%%%%%%%%%%%%%%%%%%%%%%%%%%%%%%%%%%%%%%%%

\end{document}